\documentstyle[12pt,epsfig]{article}
\begin{document}
\begin {center}
{\bf {\Large  Broadening of $\rho (\to e^+e^-)$ meson produced
              coherently in the photonuclear reaction} }
\end {center}
%\smallskip
%\medskip
\begin {center}
Swapan Das \footnote {email: swapand@barc.gov.in} \\
{\it Nuclear Physics Division,
Bhabha Atomic Research Centre  \\
Mumbai-400085, India }
\end {center}
%\medskip

\begin {abstract}
The $e^+e^-$ invariant mass distribution spectra are calculated to estimate
the hadron parameters of the $\rho$ meson produced coherently in the
photonuclear reaction.
The elementary reaction occurring in the nucleus is considered to
proceed as $ \gamma N \to \rho^0 N ; ~ \rho^0 \to e^+e^- $. We describe the
elementary $\rho$ meson photoproduction by the experimentally determined
reaction amplitude $ f_{ \gamma N \to \rho^0 N } $.
The
$\rho$ meson propagator is presented by the eikonal form, and the $\rho$
meson nucleus optical potential $V_{O\rho}$ appearing in it is estimated
using the ``$t\varrho$'' approximation. The $\rho$ meson nucleon
scattering amplitude $ f_{ \rho N } $ extracted from the measurements is
used to generate this potential.
The calculated $e^+e^-$ invariant mass distribution spectra are compared
with those measured at Jefferson Laboratory. The calculated results
for the transparency ratio are also presented.
\end {abstract}


PACS number(s): 25.20.Lj, 13.60.Le

\section{Introduction}

The vector meson production in the nuclear reaction has drawn considerable
attention to look for its properties in the nucleus
\cite{kokl,cabr,rawa,lemm,haha}.
Experimentally, large modification of the $\rho$ meson was indicated
first in the enhanced dilepton yield (between 300 and 700 MeV) in CERES
and HELIOS ultra-relativistic heavy ion collision data taken sometime
around 1995 in CERN-SPS \cite{dree,ulch,masr,scin}.
Since the statistics and resolution
of this yield was poor, the quantitative estimation of the $\rho$ meson
parameters could not be made in that time.
Theoretically, these data have been found compatible with two different
approaches:
(i) the dropping of $\rho$ meson mass \cite{lkb1,lkb2,cek1,cek2},
and (ii) the many body
interaction of $\rho$ meson with other hadrons in the nuclear medium
\cite{carw,racw,cbrw,rawm,resw}.
Almost a decade later, the STAR experiment at RHIC-BNL \cite{rhic} found
the decrease in $\rho$ meson mass $\sim 70$ MeV in the analysis of the
$\pi^+\pi^-$ production data from the peripheral Au+Au collisions. In
contrast, significant broadening in the $\rho$ meson mass distribution
spectrum, but essentially no shift in mass, was reported by the upgraded
CERES experiment \cite{uceres} and dimuon measurements (in the In-In
collision) in the NA60 experiment at CERN
\cite{ardi1,ardi2,daic}.

The in-medium properties of vector meson can be searched in the normal
nuclear density and zero temperature where the calculations and
interpretations for these quantities can be done judiciously.
The
scaling hypothesis due to Brow and Rho \cite{br} and the QCD sum rule
calculation due to Hatsuda and Lee \cite{hat} envisage the reduction of
vector meson mass in the nucleus. Saito et al. \cite{sai}, using the
quark meson coupling model, have established a relation which describes
the drop of $\rho$ and $\omega$ meson mass in the nuclear medium.
The vector meson dominance model calculations due to Asakawa et al.
\cite{aklq,ask1,ask2}
show the reduction of the $\rho$ meson mass with the increase
in nuclear density.
The modification of the vector mesons is elucidated large enough to
observe it in the nuclear reactions with photon and pion beams
\cite{effn,wbcm,ebcm}.
The broadening in the $\rho$ meson width, in addition to the upward
mass-shift and the appearance of additional peaks of this meson, in the
nuclear matter are reported in Ref.~\cite{cov}. The nuclear bound state
calculation for the $\omega$ meson \cite{mwnc} show the drop of mass
and increase in width of this meson in the hadron induced nuclear
reactions. The modification of $\phi$ meson in the nuclear matter as
well as in the proton and photon induced nuclear reactions is discussed
in Ref.~\cite{cavi,crotv,maro}.

During the last decade, extensive measurements had been pursued to
search the hadron parameters of the vector meson in the normal nucleus.
The CBELSA/TAPS collaboration measured the $\pi^0\gamma$ invariant mass
distribution spectrum to look for the $\omega$ meson parameters in Nb
nucleus \cite{elsa}. The line shape of the $\omega$ meson in this
reaction, as shown in our calculation \cite{das}, does not show medium
effect on it \cite{metag1}. Other experiments on $ ( \gamma, \omega p ) $
reaction corroborate this result \cite{nan1, nan2}.
Recent measurements on
the nuclear transparency ratio \cite{ktla, wood}
reported large in-medium
width of the $\omega$ and $\phi$
mesons.
The KEK-PS E325 collaboration at KEK \cite{mut1,nkri,mut2}
found the enhancement
in the $e^+e^-$ yield in the $p+$A reaction at 12 GeV. This enhancement
is well understood due to the reduction of the vector meson mass in the
nucleus.
TAGX collaboration \cite{huber} reported the in-medium boardening
(without mass-shift) of the $\rho$ meson produced in the
$ ( \gamma, \pi^+\pi^- ) $ reaction on $^{12}$C
nucleus.
The recent data measured by CLAS collaboration in Jefferson Laboratory
also show no mass-shift but significant broadening of the $\rho$ meson
produced in the photonuclear reaction \cite{nspr, wod2}.
In this measurement,
the $\rho^0$ meson was produced in nuclei by the tagged photon beam
of energy range $0.61-3.82$ GeV and it was probed by $e^+e^-$ for its
momentum 0.8 to 3.0 GeV/c.

To study the in-medium properties of $\rho$ meson, we calculate the
differential cross section of the $e^+e^-$ invariant mass distribution
in the coherent $\rho$ meson photoproduction reaction on nuclei. The
coherent meson production in the nuclear reaction is a potential tool
to investigate the meson dynamics in the nucleus.
In
our study, the elementary reaction occurring in the nucleus is visualized
as $ \gamma N \to \rho^0 N ; ~ \rho^0 \to e^+e^- $. The $\rho$ meson
photoproduction is described by the reaction amplitude
$ f_{\gamma N \to \rho N} $.
We
address the $\rho$ meson propagator by the eikonal form. The $\rho$ meson
nucleus interaction appearing in it is described by the corresponding
optical potential, evaluated using $``t\varrho"$ approximation. The decay
of $ \rho^0 \to e^+e^- $ is described by the corresponding decay width.
It could be mentioned that there exist other mechanisms for the $\rho$
meson production in the photonuclear reaction. We will study those
separately. The calculated $e^+e^-$ invariant mass (i.e., $\rho^0$ meson
mass) are compared with the data from Jefferson Laboratory
\cite{nspr, wod2}
to see at what extent the calculated results can reproduce the measured
spectrum.

The $\rho$ meson production in the photonuclear reaction in the energy
region of Jefferson Laboratory is also studied by Effenberger et al.,
\cite{effn} and Riek et al., \cite{rrol1, rrol2}.
The previous authors have done semiclassical transport model calculation
where they have discussed various aspect of $e^+e^-$ emission in the
photonuclear reaction.
For
the $\rho$ meson production, they have accounted it by the parameterization
of the elementary $\rho$ meson photoproduction (on a nucleon) data. The
medium effect is evaluated though the collision broadening and the
potential extracted from the mass-shift formulated by the Hatsuda and Lee
\cite{hat}. This potential is multiplied by a factor to incorporate the
momentum dependence for it.
The
later authors have described the elementary $\rho$ meson photoproduction
by the meson/pomeron exchange at $ E_\gamma \geq 2 $ GeV. At lower energy,
they have described it by the $s$ wave resonances whose parameters are
taken from the in-medium $\rho$ meson self-energy appearing in its spectral
function. In fact, this self-energy (calculated in the hadronic many
body theory) is used to estimate the medium effect on the $\rho$
meson.
We describe the $\rho$ meson photoproduction and the $\rho$ meson potential
(which illustrates the medium effect on this meson) by its scattering
amplitude. The energy dependent values for this amplitude is extracted
from the elementary $ \gamma N \to \rho N $ reaction data. Therefore,
we use (unlike others) the experimentally determined inputs in the
calculation.

\section{Formalism}

The production of $\rho$ meson in the photonuclear reaction can be
described by the generalized optical potential or self-energy of this
meson  \cite{das, pash} as
\begin{equation}
\Pi_{\gamma A \to \rho A} ( {\bf r} ) =
-4\pi E_\rho \left [ \frac{1}{{\tilde E}_\rho} +
\frac{1}{{\tilde E}_N} \right ]
f_{\gamma N \to \rho N} (0) \varrho ({\bf r}),
\label{gpa}
\end{equation}
where ${\tilde E}_\rho$ and ${\tilde E}_N$ are the energies of the
$\rho$ meson and nucleon respectively in the $\rho N$  c.m. system of
energy equal to $\gamma N$ c.m. energy. $ f_{\gamma N \to \rho N} (0) $
denotes the amplitude for the $\gamma N \to \rho N$ reaction, and
$\varrho ({\bf r})$ represents the spatial density distribution of
the nucleus. These quantities are described elaborately in the next
section.

The propagation of the $\rho$ meson from its production point ${\bf r}$ to
its decay point ${\bf r^\prime}$ can be expressed as
$ ( -g^\mu_{\mu^\prime} + \frac{1}{m^2} k^\mu_\rho k_{\rho,\mu^\prime} )
G_\rho ( m; {\bf r^\prime - r} ) $ \cite{das, das3}. It is shown later the
emission of $\rho$ meson of 0.8 to 3.0 GeV/c momentum is strongly
focused towards the forward direction. This occurs since the kinetic energy
of the $\rho$ meson is much larger than its potential energy.
We, therefore, represent the scalar part of the $\rho$ meson propagator
$ G_\rho ( m; {\bf r^\prime - r} ) $ by the eikonal form \cite{das5, gkc},
i.e.,
\begin{equation}
G_\rho ( m; {\bf r^\prime - r} ) = \delta ( {\bf b^\prime - b} )
\Theta (z^\prime - z) e^{ i {\bf k_\rho \cdot (r^\prime - r)} }
D_{\bf k_\rho} ( m; {\bf b}, z^\prime, z ).
\label{omp}
\end {equation}
The factor $ D_{\bf k_\rho} ( m; {\bf b}, z^\prime, z ) $ in this equation
describes the nuclear medium effect on the properties of $\rho$ meson. The
form for it is
\begin{equation}
D_{\bf k_\rho} ( m; {\bf b}, z^\prime, z ) =
-\frac{i}{ 2k_{\rho\parallel} }
exp \left [  \frac{i}{ 2k_{\rho\parallel} } \int_z^{z^\prime}
 dz^{\prime \prime} \{ {\tilde G}^{-1}_{0\rho} ( m ) -
    2 E_\rho V_{O\rho} ({\bf b}, z^{\prime \prime}) \} \right  ],
\label{dom}
\end{equation}
where $k_\rho$ is the momentum of the $\rho$ meson.
$ V_{O\rho} ({\bf b}, z^{\prime \prime}) $ represents the $\rho$ meson
nucleus optical potential. This potential can modify the hadronic
parameters of the $\rho$ meson during its passage through the nucleus.
$ {\tilde G}_{0\rho} (m) $ denotes the $\rho$ meson (on-shell)
propagator in free space: $ {\tilde G}^{-1}_{0\rho} (m)
= m^2-m^2_\rho+im_\rho\Gamma_\rho (m) $. Here, $m_\rho$ and
$\Gamma_\rho (m)$ represent the resonant mass and total decay width
respectively of the $\rho$ meson.

The differential cross section for the dilepton invariant mass $m$
(arising due to the decay of $\rho^0$ meson of mass $m$) distribution
in the coherent $ ( \gamma, \rho^0 \to e^+e^- ) $ reaction on a nucleus
can be written as
\begin{equation}
\frac{ d\sigma (E_\gamma) }{ dm }
= \int d\Omega_\rho K_F \Gamma_{\rho^0 \to e^+ e^-} (m)
                   | F({\bf k}_\gamma, {\bf k}_\rho) |^2,
\label{dsc1}
\end{equation}
where $ \Gamma_{\rho^0 \to e^+e^-} (m) $ stands for the width of the
$\rho$ meson of mass $m$ decaying at rest into dilepton:
$ \Gamma_{\rho^0 \to e^+e^-} (m) \approx  8.8 \times 10^{-6} m $
\cite{shmt,bdri,skri}.
$K_F$ denotes the kinematical factor associated in the reaction. It is
given by
\begin{equation}
K_F = \frac{3\pi}{(2\pi)^4} \frac{ k^2_\rho E_{A^\prime} m^2 }
      { E_\gamma | k_\rho E_i - {\bf k}_\gamma \cdot {\hat k}_\rho E_\rho | } ;
\label{kfc1}
\end{equation}
with $ {\bf k}_\rho = {\bf k}_{e^+} + {\bf k}_{e^-} $. $E_{A^\prime}$ is
the energy of the recoil nucleus. All other symbols carry their usual
meanings.

The factor $ F ({\bf k}_\gamma, {\bf k}_\rho) $ in Eq.~(\ref{dsc1})
describes the photoproduction of $\rho^0$ meson in the nucleus as well
as the propagation of this meson through the nucleus. The expression
for $ F ({\bf k}_\gamma, {\bf k}_\rho) $ is
\begin{equation}
F ({\bf k}_\gamma, {\bf k}_\rho)
= \int d{\bf r} \Pi_{\gamma A \to \rho A} ({\bf r})
   e^{ i({\bf k}_\gamma - {\bf k}_\rho) . {\bf r} }
   D ({\bf k}_\rho; {\bf b}, z),
\label{fdfn}
\end{equation}
where $ D ({\bf k}_\rho; {\bf b}, z) $ is given by
\begin{equation}
D ({\bf k}_\rho; {\bf b}, z)
= \int^\infty_z dz^\prime D_{\bf k_\rho} (m; {\bf b}, z^\prime, z).
\label{ybio}
\end{equation}
It should be mentioned that $ F ({\bf k}_\gamma, {\bf k}_\rho) $ carries
the information about the in-medium properties of the $\rho$ meson, see
in Eq.~(\ref{dom}).

The Eq.~(\ref{dsc1}) can be used to estimate the differential cross
section for the $ \rho^0 (\to e^+e^-) $ meson mass distribution due
to fixed $\gamma$ beam energy $E_\gamma$. As mentioned earlier, the
tagged photon beam was used in the measurements done at Jefferson
Laboratory \cite{nspr, wod2}.
The energies of this beam are weighted in 6
bins to simulate the beam profile used by CLAS collaboration
\cite{rrol1} (also see the reference there in). Therefore, the cross
section can be expressed as
\begin{equation}
\frac{d\sigma}{dm} = \sum^6_{i=1}
W(E_{\gamma,i}) \frac{d\sigma (E_{{\gamma,i}})} {dm},
\label{dsc2}
\end{equation}
where $d\sigma(E_{\gamma,i}) / dm$ is given in Eq.~(\ref{dsc1}).
$E_{\gamma,i}$ consists of six bins of photon energies,
$E_{\gamma,i}$ (GeV) = 1.0, 1.5, 2.0, 2.5, 3.0 and 3.5 with relative
weights $W(E_{\gamma,i})$ of $13.7\%$, $23.5\%$, $19.3\%$, $20.1\%$,
$12.6\%$ and $10.9\%$ respectively \cite{rrol2}.

\section{Results and Discussions}

We calculate the differential cross sections for the dilepton invariant
mass, i.e., the $\rho^0$ meson mass $m$, distribution in the coherent
$ (\gamma, \rho^0 \to e^+e^-) $ reaction on $^{12}$C, $^{48}$Ti, $^{56}$Fe
and $^{208}$Pb nuclei.
Since this meson, as mentioned earlier, is detected for its momentum
range 0.8 to 3.0 GeV/c \cite{nspr, wod2},
we imposed this constrain in the
calculation. The $\rho^0$ meson nucleus optical potential, i.e.,
$V_{O\rho}$ appearing in Eq.~(\ref{dom}), is estimated using the
$``t\varrho"$ approximation \cite{das5, eiko, glbr}:
\begin{equation}
V_{O\rho} ({\bf r}) =
-\frac{v_\rho}{2} [i+\alpha_{\rho N}] \sigma_t^{\rho N} \varrho ({\bf r}),
\label{opt}
\end{equation}
where $v_\rho$ is the velocity of the $\rho$ meson. $ \varrho ({\bf r}) $,
the nuclear density distribution, for $^{12}$C nucleus can be written as
\begin{eqnarray}
\varrho ({\bf r})
= \varrho_0 ( 1 + w r^2/c^2 ) e^{-r^2/c^2},
\label{rhr1}
\end{eqnarray}
with $w=1.247$, $c=1.649$ fm \cite{andt}. For $^{48}$Ti, $^{56}$Fe and
$^{208}$Pb nuclei, $ \varrho ({\bf r}) $ is given by
\begin{eqnarray}
\varrho ({\bf r}) = \frac{ \varrho_0 }{ 1 + e^{(r-c)/z} },
\label{rhr2}
\end{eqnarray}
where $c$ is equal to 3.754 fm for $^{48}$Ti, 3.971 fm for $^{56}$Fe and
6.624 fm for $^{208}$Pb. The value of $z$ (in fm) is equal to 0.567 for
$^{48}$Ti, 0.5935 for $^{56}$Fe and 0.549 for $^{208}$Pb \cite{andt}.
The above forms of density distribution are extracted from the electron
scattering data, and they are normalized to the mass number of the
corresponding nucleus.

\begin{figure}[h]
%\begin{figure}
\begin{center}
\centerline {\vbox {
%\psdraft
\psfig{figure=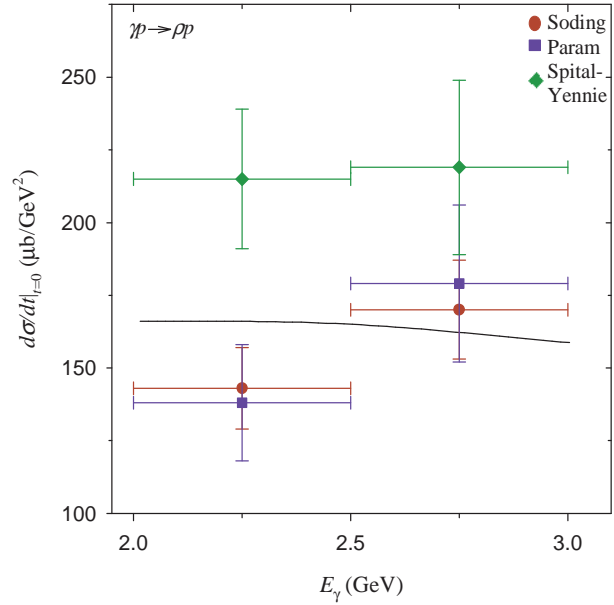,height=08.0 cm,width=08.0 cm}
}}
\caption{
(color online).
The calculated forward four-momentum transfer
$ \frac{ d\sigma }{ dt } |_{t=0}$ in the $\gamma p \to \rho p$ reaction
at various $\gamma$ energies. The data points are those extracted from
the measured $ \frac{ d\sigma }{ dt } $ distribution: solid circle
due to S$\ddot{\mbox{o}}$ding model; solid rectangle due to parametric
fit and solid diamond due to Spital-Yennie procedure (see the detail in
Ref.~\cite{bsy}).
}
\label{fgfmt}
\end{center}
\end{figure}

The scattering parameters  $\alpha_{\rho N}$ and $\sigma_t^{\rho N}$
in Eq.~(\ref{opt}) represent the ratio of the real to imaginary part
of the $\rho^0$ meson nucleon scattering amplitude $f_{\rho N}$, and
the corresponding total cross section respectively. Since the $\rho$
meson is a short lived ($\sim 10^{-23}$ second) particle, $f_{\rho N}$
can't be obtained directly from measurements.
But
this amplitude for the $\rho$ meson momentum $ k_\rho \ge 0.8 $ GeV/c is
extracted from the elementary $\rho$ meson photoproduction
($ \gamma N \to \rho N $) data \cite{kscge}.
We use this amplitude to estimate the forward
four-momentum transfer in the $ \gamma N \to \rho N $ reaction, i.e.,
$ \frac{d\sigma}{dt} (\gamma N \to \rho N) |_{t=0} $ \cite{kscge}:
\begin{equation}
\frac{d\sigma}{dt} (\gamma N \to \rho N) |_{t=0} =
\frac{\alpha_{em}}{16\gamma_\rho^2}
\left ( \frac{\tilde k_\rho}{\tilde k_\gamma} \right )^2
[1+\alpha^2_{\rho N}] (\sigma_t^{\rho N})^2,
\label{frmt}
\end{equation}
where $ \alpha_{em} (=1/137.04) $ is the fine structure constant and
the value of $\gamma_\rho$ is 2.52.
$\tilde k_\rho$ and $\tilde k_\gamma$ are the c.m. momenta of the
$\rho N$ and $\gamma N$ systems at the same c.m. enery.
The calculated $ \frac{d\sigma}{dt} (\gamma N \to \rho N) |_{t=0} $ is
presented in Fig.~\ref{fgfmt} along with that extracted from the measured
$ \frac{d\sigma}{dt} $ in the $ \gamma N \to \rho N $ reaction. The later
arises (as discussed elaborately in Ref.~\cite{bsy}) due to
S$\ddot{\mbox{o}}$ding model (solid circle), parametric fit
(solid rectangle) and Spital-Yennie procedure (solid diamond). This figure
shows that $\frac{d\sigma}{dt} (\gamma N \to \rho N) |_{t=0}$, calculated
for $2-3$ GeV of beam energy, are within the range of measured
values.
It could be mentioned that the vector meson dominance model relates
the reaction amplitude $f_{\gamma N \to \rho N}$ in Eq.~(\ref{gpa}) to
$ f_{\rho N} $ as $ f_{\gamma N \to \rho N}
= \frac{\sqrt{\pi \alpha_{em}}}{\gamma_\rho} f_{\rho N} $ \cite{kscge}.
The quantity $|f_{\gamma N \to \rho N}|^2$, as required to calculate the
cross section in few GeV regoin, can be extracted directly from the
measured differential cross section \cite{stlr}:
$ \frac{d\sigma}{dt} (\gamma N \to \rho N) |_{t=0}
\simeq \frac{ \pi }{ \tilde {k}^2_\gamma } |f_{\gamma N \to \rho N}|^2$.

\begin{figure}[h]
%\begin{figure}
\begin{center}
\centerline {\vbox {
%\psdraft
\psfig{figure=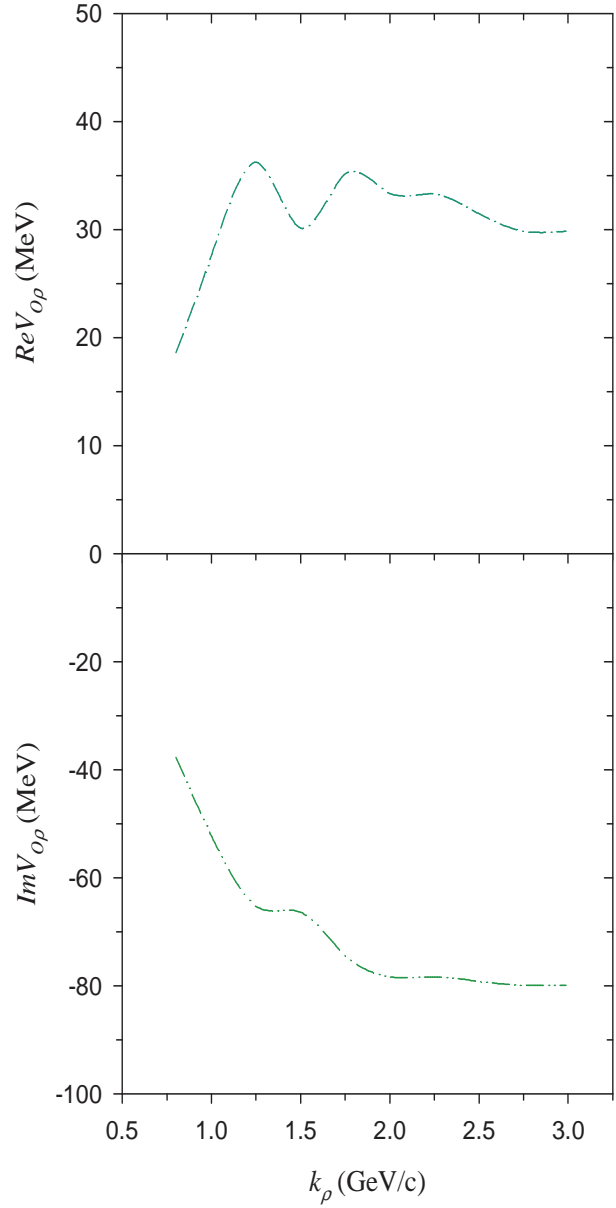,height=16.0 cm,width=08.0 cm}
}}
\caption{
(color online).
The variation of the $\rho^0$ meson optical potential $V_{O\rho}$ with
its momentum $k_\rho$. $Re V_{O\rho}$ and $Im V_{O\rho}$ represent the
real and imaginary parts of $V_{O\rho}$ respectively.
This potential is calculated for the central density of $^{12}$C nucleus.
}
\label{fgopt}
\end{center}
\end{figure}

\begin{figure}[h]
%\begin{figure}
\begin{center}
\centerline {\vbox {
%\psdraft
\psfig{figure=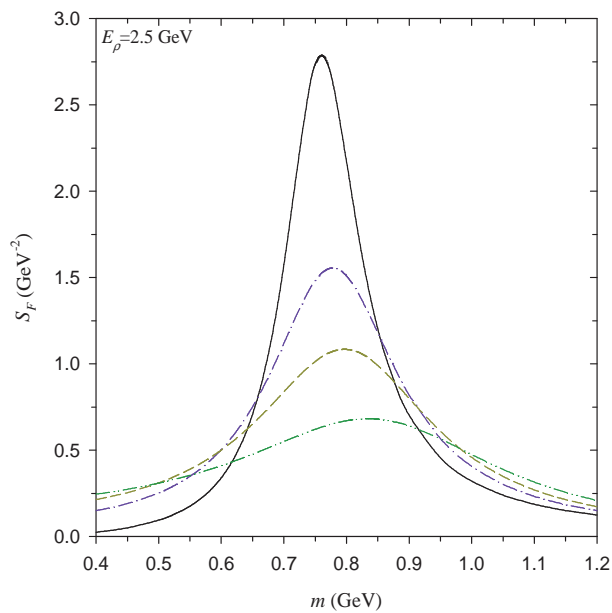,height=08.0 cm,width=08.0 cm}
}}
\caption{
(color online).
The $\rho$ meson spectral function $S_F$ at $E_\rho \simeq 2.5$ GeV
resulting from the optical potential $V_{O\rho}$ in Eq.~(\ref{opt}) for
several nuclear densities $\varrho$. The nuclear saturation density
$\varrho_s$ is taken equal to 0.17 fm$^{-3}$. The solid, dot-dashed,
dashed and dot-dot-dashed lines are due to densities: $\varrho$ = 0,
$\varrho_s/4$, $\varrho_s/2$ and $\varrho_s$ respectively.
}
\label{fsfn}
\end{center}
\end{figure}

The $\rho^0$ meson optical potential $V_{O\rho} ({\bf r}=0)$ in
Eq.~(\ref{opt}), evaluated using the $\rho$ meson nucleon scattering
parameters, is presented in Fig.~\ref{fgopt} for $^{12}$C nucleus.
This potential can modify the in-medium $\rho$ meson spectral function
$S_F$, defined as
\begin{equation}
S_F = -\frac{1}{\pi} Im \left [
 \frac{1}{m^2-m^2_\rho+im_\rho\Gamma_\rho(m)-2E_\rho V_{O\rho}} \right ].
\label{spfn}
\end{equation}
The behavior of $S_F$ at $ E_\rho \simeq 2.5 $ GeV is presented in
Fig.~\ref{fsfn} for several nuclear densities, including that in the free
space. This figure illustrates that the spectral function $S_F$ broadens
and its peak position shifts towards the higher values, as the density
$\varrho$ of the nucleus increases. Compare to free space $S_F$, the
width of the in-medium $S_F$ is enhanced from 146 MeV to 546 MeV and its
peak position is shifted from 760 MeV to 830 MeV due to the $\rho$ meson
potential $V_{O\rho}$ resulting from the nuclear saturation density
$\varrho_s$, i.e., 0.17 fm$^{-3}$.

\begin{figure}[h]
%\begin{figure}
\begin{center}
\centerline {\vbox {
%\psdraft
\psfig{figure=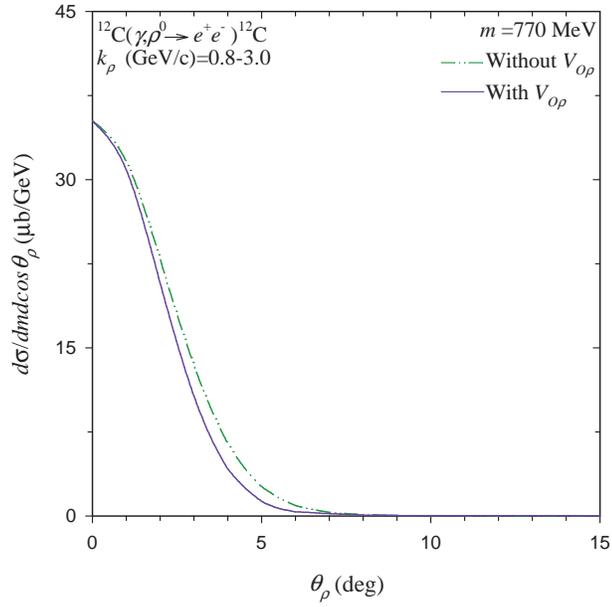,height=08.0 cm,width=08.0 cm}
}}
\caption{
(color online).
The angular distributions of the $\rho^0$ meson, photoproduced in the
$^{12}$C nucleus for $ k_\rho = 0.3 -3.0 $ GeV/c, calculated with and
without the $\rho$ meson optical potential $V_{O\rho}$. The cross section
at the peak is reduced by a factor of 2.55 due to
$V_{O\rho}$. The $\rho$ meson mass $m$ is taken equal to 770 MeV.
}
\label{fgan}
\end{center}
\end{figure}

The calculated angular distribution $ d\sigma / dm dcos\theta_\rho $
of the $\rho^0$ meson photoproduced in the $^{12}$C nucleus is presented
in Fig.~\ref{fgan}. The $\rho$ meson mass $m$ is taken equal to 770
MeV.
The dot-dot-dashed curve in this figure refers to the $\rho$ meson angular
distribution for its free propagation through the nucleus, i.e., the $\rho$
meson nucleus interaction $V_{O\rho}$ is not included in the calculated
results. The solid curve represents that for $V_{O\rho}$ incorporated in
the calculated angular distribution.
The peak cross section is reduced by a factor of 2.55 (i.e., from
$\sim$35.25 $\mu$b/GeV to 13.81 $\mu$b/GeV) due to $V_{O\rho}$, where as
the change in the shape of the angular distribution due to it is negligible.
As shown in this figure, the $\rho$ meson of momentum $0.8-3.0$ GeV/c
is emitted distinctly in the forward direction.

\begin{figure}[h]
%\begin{figure}
\begin{center}
\centerline {\vbox {
%\psdraft
\psfig{figure=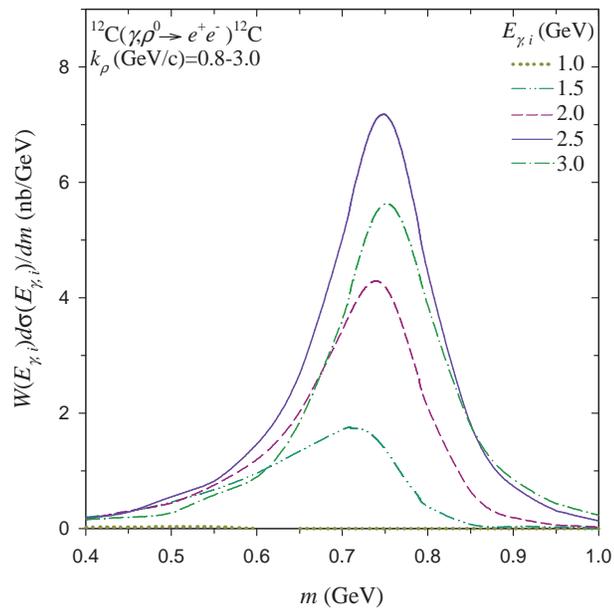,height=08.0 cm,width=08.0 cm}
}}
\caption{
(color online).
The calculated cross section for the $\rho^0$ meson mass $m$ distribution
at various beam energies. The cross section is maximum at 2.5 GeV.
}
\label{fgbm}
\end{center}
\end{figure}

In Eq.~(\ref{dsc2}), $ W( E_{\gamma, i} ) d\sigma ( E_{\gamma, i} ) /dm $
represents the cross section of the $\rho^0$ meson mass distribution at
the beam energy $ E_{\gamma, i} $. We plot this cross section for various
$ E_{\gamma, i} $ in Fig.~\ref{fgbm}.
The solid curve in this figure shows the cross section is maximum, i.e.,
7.18 nb/GeV, at 2.5 GeV beam energy. The cross sections for beam energies
equal to 2.0 GeV (dashed curve) and 3.0 GeV (dot-dashed curve) are also
significant.
At 3.5 GeV, the $\rho$ meson momentum is larger than 3.0 GeV/c. Therefore,
the cross section is not calculated for this energy because of the
kinematical restriction, i.e., $ k_\rho = 0.8 - 3.0 $ GeV/c, imposed
on the $\rho$ meson momentum.

\begin{figure}[h]
%\begin{figure}
\begin{center}
\centerline {\vbox {
%\psdraft
\psfig{figure=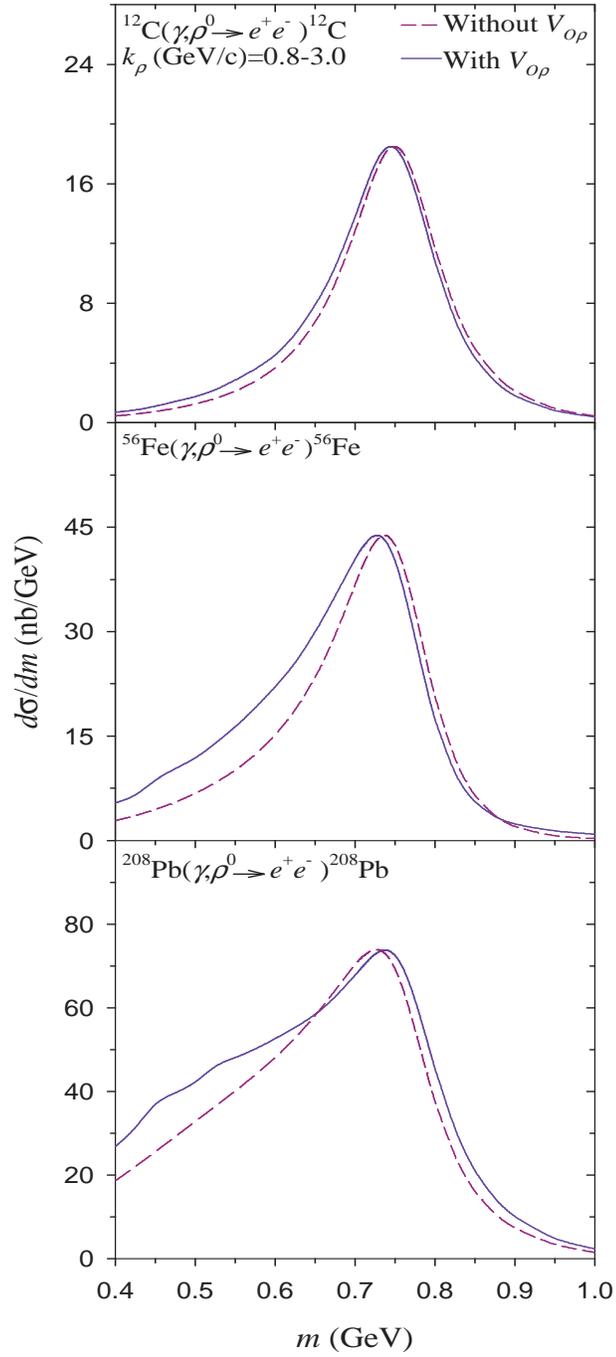,height=18.0 cm,width=08.0 cm}
}}
\caption{
(color online).
The calculated cross sections for the $\rho$ meson mass distribution
with and without the $\rho^0$ meson nucleus interaction $V_{O\rho}$
(see text).
The peak cross section is reduced due to $V_{O\rho}$ by the factor of
3.06 for $^{12}$C, 4.5 for $^{56}$Fe and 3.55 for $^{208}$Pb nuclei.
}
\label{fgmd}
\end{center}
\end{figure}

The $\rho$ meson parameters can be modified in the nuclear reaction
due to its interaction with the nucleus, i.e., $V_{O\rho}$ in
Eq.~(\ref{opt}). Therefore, the calculated results with and without
$V_{O\rho}$ can elucidate the nuclear medium effect on the $\rho$ meson.
The dashed curves in Fig.~\ref{fgmd} represent the calculated $\rho$
meson mass distribution spectra without this interaction. The solid curves
in this figure represent those due to the inclusion of $V_{O\rho}$ in the
calculation.
The cross sections are attenuated due to $V_{O\rho}$ from 56.42 nb/GeV
to 18.43 nb/GeV for $^{12}$C, from 196.88 nb/GeV to 43.78 nb/GeV for
$^{56}$Fe and from 262.12 nb/GeV to 73.78 nb/GeV for $^{208}$Pb at the
respective peaks
which appear at the $\rho$ meson mass around 740 MeV. The shift in the
peak position due to this potential is insignificant, i.e., $\leq$10
MeV, for all of these nuclei.
The calculated results in Fig.~\ref{fgmd} also show that the width of the
$\rho$ meson mass distribution spectrum increases with the size of the
nucleus due to the $\rho$ meson nucleus interaction $V_{O\rho}$. The
enhancement in the width is found equal to 6 MeV for $^{12}$C, 35
MeV for $^{56}$Fe, and 92 MeV for $^{208}$Pb.

The change in the shape of $\rho$ meson mass distribution spectrum
occurs due to the decay of this meson inside the nucleus. For
illustration \cite{gkc}, we consider the $\rho$ meson decay within the
nuclear density $ \varrho (r) \leq 0.01 \varrho (0) $ as the $\rho$
meson decay inside the nucleus. The mass distribution of the $\rho$ meson
decaying inside and outside the nucleus can be studied by splitting
$ D ({\bf k}_\rho; {\bf b}, z) $ in Eq.~(\ref{ybio}) into two parts
\cite{gkc}:
\begin{equation}
D ({\bf k}_\rho; {\bf b}, z)
= D_{in} ({\bf k}_\rho; {\bf b}, z) + D_{out} ({\bf k}_\rho; {\bf b}, z),
\label{ybio2}
\end{equation}
where $ D_{in} ({\bf k}_\rho; {\bf b}, z) $ and
$ D_{out} ({\bf k}_\rho; {\bf b}, z) $ describe the $\rho$ meson
decay inside and outside the nucleus respectively. Using Eq.~(\ref{ybio}),
they can be written as
\begin{eqnarray}
&D_{in} ({\bf k}_\rho; {\bf b}, z)&
= \int^Z_z dz^\prime  D_{\bf k_\rho} ({\bf b}, z^\prime, z),    \\
\label{ybi}
&D_{out} ({\bf k}_\rho; {\bf b}, z)&
= \int^\infty_Z dz^\prime  D_{\bf k_\rho} ({\bf b}, z^\prime, z),
\label{ybo}
\end{eqnarray}
with $ Z=\sqrt{R^2-b^2} $. $R$ is the radius of the nucleus where its
density falls to $1\%$ of the saturation density.

\begin{figure}[h]
%\begin{figure}
\begin{center}
\centerline {\vbox {
%\psdraft
\psfig{figure=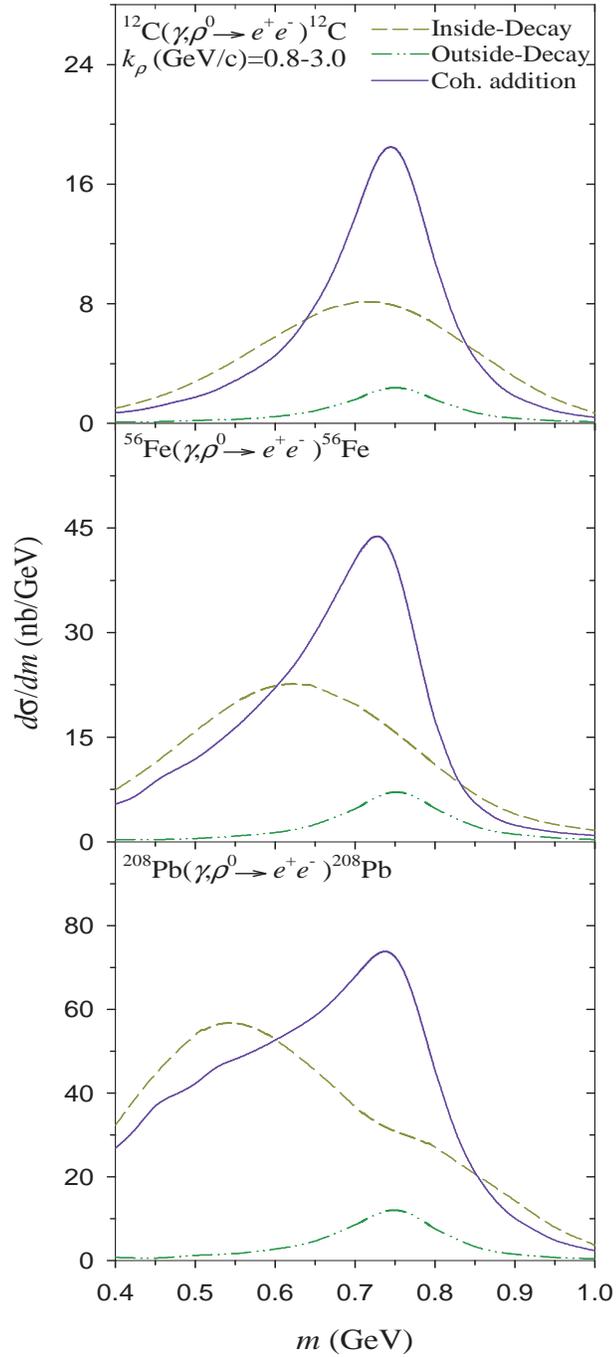,height=18.0 cm,width=08.0 cm}
}}
\caption{
(color online).
The dashed curves represent the calculated mass distribution spectra
for the $\rho^0$ meson decaying inside the nucleus where as the
dot-dot-dashed curves denote those for the $\rho^0$ meson decaying
outside the nucleus. The coherent addition of them is shown by the solid
curves.
}
\label{fgdy}
\end{center}
\end{figure}

The dashed curves in Fig.~\ref{fgdy} distinctly elucidate that the
$\rho$ meson mass distribution becomes wider and the peak shifts
towards the lower value when it decays inside the larger
nucleus. The width (in MeV) of the $\rho^0$ meson decaying inside
$^{12}$C, $^{56}$Fe and $^{208}$Pb nuclei are 327.69, 347.68 and
401.53 respectively. The corresponding peak appears at 710, 620 and
540 MeV.
For the $\rho$ meson decaying outside the nucleus, its mass distribution
spectra (shown by the dot-dot-dashed curves) do not change with the
nucleus.
These distributions show the peak appears at 750 MeV and the width is
$\sim$150 MeV for all nuclei. The solid curves represent the cross
section due to the coherent addition of the amplitudes of the $\rho$
meson decaying inside and outside the nucleus.

\begin{figure}[h]
%\begin{figure}
\begin{center}
\centerline {\vbox {
%\psdraft
\psfig{figure=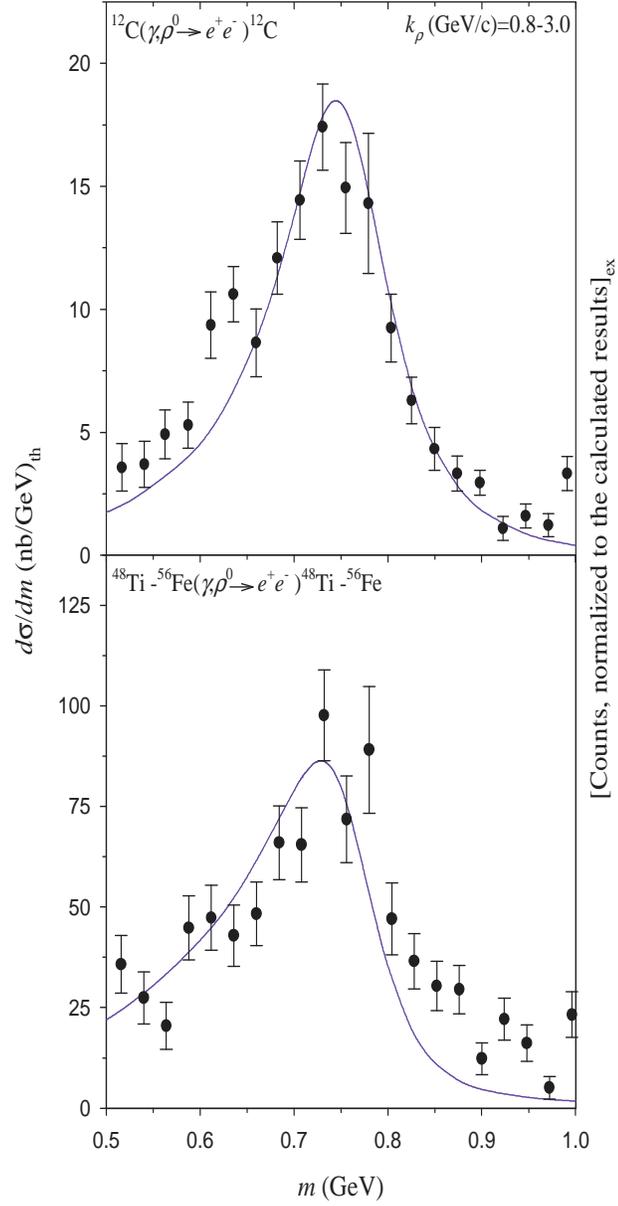,height=16.0 cm,width=08.0 cm}
}}
\caption{
(color online).
The calculated $\rho^0$ meson mass distribution spectra (solid curves)
are compared with the data, taken from Ref.~\cite{nspr, wod2}.
}
\label{fgtd}
\end{center}
\end{figure}

The calculated $\rho^0$ meson mass distribution spectra $d\sigma/dm$
(solid curve) are presented in Fig.~\ref{fgtd} along with the measured
$e^+e^-$ invariant mass distribution spectra \cite{nspr, wod2},
since the
dilepton pair $e^+e^-$ arises (as mentioned earlier) due to the
decay of the $\rho^0$ meson photoproduced in the nucleus.
We
compare the calculated results with the data to investigate at what extent
these results can reproduce the shape of the measured spectra. The upper
part of this figure shows the calculated and measured spectra for
$^{12}$C nucleus,
where as those for $^{48}$Ti and $^{56}$Fe nuclei are shown in the lower
part of it. For these nuclei, we add the cross sections calculated
separately and show them with the data. As shown in this figure, the
calculated spectra reproduce reasonably the shape of the measured
distributions.

The nuclear transparency ratio $T_A$ is another way to look for the
nuclear effect on the $\rho$ meson propagating through the nucleus.
It is the ratio of the cross section per nucleon in a nucleus relative
to that of the free nucleon. Hence, $T_A$ measures the suppression of
the elementary cross section in the nucleus. As discussed in
Ref.~\cite{rrol2}, we can write
\begin{equation}
T_A \simeq \frac{ \int dm (d\sigma/dm)_{med} }{ \int dm (d\sigma/dm)_{vac} }.
\label{trns}
\end{equation}
The values for $T_A$ are 0.34 for $^{12}$C, 0.26 for $^{56}$Fe and 0.32
for $^{208}$Pb nuclei. It shows anomalous result for Pb nucleus.

The calculated results showing $e^+e^-$ emission in the photonuclear
reaction (in the energy region of Jefferson Laboratory) also have been
reported by Effenberger et al., \cite{effn} and Riek et al.,
\cite{rrol1, rrol2}.
The previous authors have shown an additional peak in the dilepton yield
at 650 MeV due to the mass modification of the vector mesons, which gets
washed out because of the inclusion of the collision broadening. With the
use of the momentum-dependent potential they have obtained the results
close to that of the bare mass case.
The
later authors have studied $e^+e^-$ emission due to the decay of the
$\rho$ meson, and their results reproduce well the measured spectrum
reported from Jefferson Laboratory  \cite{nspr, wod2}.
They have also shown
systematically the enhancement in the $\rho$ meson width with the nuclear
density as well as with the size of the nucleus.
We also consider the $\rho$ meson as the source of $e^+e^-$ emission.
Our results show the broadening of the $\rho$ meson with the size of the
nucleus, and reproduce reasonably the data from Jefferson Laboratory
\cite{nspr, wod2}.
For the calculation of the transparency ratio, we get consistent results
for C and Fe nuclei but obtain anomalous result for Pb nucleus where as
Riek et al., \cite{rrol2} have shown consistent results for all nuclei.

\section{Conclusions}

We have studied the mechanism for the coherent
$( \gamma, \rho^0 \to e^+e^- )$ reaction on the nucleus, and calculated
the differential cross sections for the $e^+e^-$ invariant mass
(i.e., $\rho^0$ meson mass) distribution in this reaction.
The reduction in the cross section as well as the enhancement in the width
(specifically in the heavy nucleus) of the $\rho$ meson mass distribution
spectrum occur due to the $\rho$ meson nucleus interaction. The peak-shift
of the calculated spectrum due to this interaction is found
insignificant.
The modification of the $\rho$ meson width arises due to the decay
of this meson inside the nucleus, which increases with the size of the
nucleus. The calculated spectra are accord with the data. However, the
calculated transparency ratio shows anomalous result for Pb nucleus.

\section{Acknowledgement}

The author thanks Prof. R. Nasseripour for sending the data for Fe-Ti
nuclei.

\newpage

{\bf Figure Captions}

\begin{enumerate}

\item (color online).
The calculated forward four-momentum transfer
$ \frac{ d\sigma }{ dt } |_{t=0}$ in the $\gamma p \to \rho p$ reaction
at various $\gamma$ energies. The data points are those extracted from
the measured $ \frac{ d\sigma }{ dt } $ distribution: solid circle
due to S$\ddot{\mbox{o}}$ding model; solid rectangle due to parametric
fit and solid diamond due to Spital-Yennie procedure (see the detail in
Ref.~\cite{bsy}).

\item (color online).
The variation of the $\rho^0$ meson optical potential $V_{O\rho}$ with
its momentum $k_\rho$. $Re V_{O\rho}$ and $Im V_{O\rho}$ represent the
real and imaginary parts of $V_{O\rho}$ respectively.
This potential is calculated for the central density of $^{12}$C nucleus.

\item (color online).
The $\rho$ meson spectral function $S_F$ at $E_\rho \simeq 2.5$ GeV
resulting from the optical potential $V_{O\rho}$ in Eq.~(\ref{opt}) for
several nuclear densities $\varrho$. The nuclear saturation density
$\varrho_s$ is taken equal to 0.17 fm$^{-3}$. The solid, dot-dashed,
dashed and dot-dot-dashed lines are due to densities: $\varrho$ = 0,
$\varrho_s/4$, $\varrho_s/2$ and $\varrho_s$ respectively.

\item (color online).
The angular distributions of the $\rho^0$ meson, photoproduced in the
$^{12}$C nucleus for $ k_\rho = 0.3 -3.0 $ GeV/c, calculated with and
without the $\rho$ meson optical potential $V_{O\rho}$. The cross section
at the peak is reduced by a factor of 2.55 due to
$V_{O\rho}$. The $\rho$ meson mass $m$ is taken equal to 770 MeV.

\item (color online).
The calculated cross section for the $\rho^0$ meson mass $m$ distribution
at various beam energies. The cross section is maximum at 2.5 GeV.

\item (color online).
The calculated cross sections for the $\rho$ meson mass distribution
with and without the $\rho^0$ meson nucleus interaction $V_{O\rho}$
(see text).
The peak cross section is reduced due to $V_{O\rho}$ by the factor of
3.06 for $^{12}$C, 4.5 for $^{56}$Fe and 3.55 for $^{208}$Pb nuclei.

\item (color online).
The dashed curves represent the calculated mass distribution spectra
for the $\rho^0$ meson decaying inside the nucleus where as the
dot-dot-dashed curves denote those for the $\rho^0$ meson decaying
outside the nucleus. The coherent addition of them is shown by the solid
curves.

\item (color online).
The calculated $\rho^0$ meson mass distribution spectra (solid curves)
are compared with the data, taken from Ref.~\cite{nspr, wod2}.

\end{enumerate}

\newpage

%\vspace{1 cm}

%\vspace{1 cm}

%\vspace{1 cm}

%\vspace{1 cm}

%\vspace{1 cm}

%\vspace{1 cm}

%\vspace{1 cm}

%\vspace{1 cm}

\end{document}